\documentclass{article}

\usepackage{arxiv}

\usepackage[utf8]{inputenc} 
\usepackage[T1]{fontenc}    
\usepackage{hyperref}       
\usepackage{url}            
\usepackage{booktabs}       
\usepackage{amsfonts}       
\usepackage{nicefrac}       
\usepackage{microtype}      
\usepackage{lipsum}
\usepackage{graphics} 
\usepackage{epsfig} 
\usepackage{times} 

\usepackage{amsmath} 
\usepackage{amssymb}  
\usepackage{amsthm}
\usepackage{amstext}
\usepackage[dvipsnames]{xcolor}
\usepackage{siunitx}
\graphicspath{{./images/}}
\usepackage{bbding}
\usepackage{amsfonts}
\newtheorem{definition}{Definition}
\newtheorem{problem}{Problem}
\newtheorem{theorem}{Theorem}
\newtheorem{remark}{Remark}
\newtheorem{assumption}{Assumption}
\newtheorem{lemma}{Lemma}
\usepackage{algorithm}
\usepackage[noend]{algpseudocode}
\DeclareMathAlphabet{\mathbbold}{U}{bbold}{m}{n}
\algrenewcommand\algorithmicrequire{\textbf{Input:}}
\algrenewcommand\algorithmicensure{\textbf{Output:}}

\usepackage{subcaption}
\usepackage{graphicx}
\usepackage{textcomp}
\usepackage{xcolor}
\usepackage{svg}
\usepackage{booktabs}

\algdef{SE}[SUBALG]{Indent}{EndIndent}{}{\algorithmicend\ }%
\algtext*{Indent}
\algtext*{EndIndent}

\title{\LARGE \bf
Neural Control Barrier Functions for Signal Temporal Logic Specifications with Input Constraints
\thanks{This work was supported by Siemens fellowship.}
}

\author{
 Vaishnavi Jagabathula\\
  Centre for Cyber-Physical Systems\\
  IISc, Bengaluru, India\\
  \texttt{vaishnavij@iisc.ac.in} \\
  \And
 Pushpak Jagtap \\
  Centre for Cyber-Physical Systems\\
  IISc, Bengaluru, India\\
  \texttt{pushpak@iisc.ac.in} \\
  }

\begin{document}
\maketitle
\thispagestyle{empty}
\pagestyle{empty}
\begin{abstract}

Signal Temporal Logic (STL) provides a powerful framework to describe complex tasks involving temporal and logical behavior in dynamical systems. This work addresses controller synthesis for continuous-time systems subject to STL specifications and input constraints. We propose a neural network-based framework for synthesizing time-varying control barrier functions (TVCBF) and their corresponding controllers for systems to fulfill a fragment of STL specifications while respecting input constraints. We formulate barrier conditions incorporating the spatial and temporal logic of the given STL specification. We also incorporate a method to refine the time-varying set that satisfies the STL specification for the given input constraints. Additionally, we introduce a validity condition to provide formal safety guarantees across the entire state space. Finally, we demonstrate the effectiveness of the proposed approach through several simulation studies considering different STL tasks for various dynamical systems (including affine and non-affine systems). 
\end{abstract}



\maketitle


\section{Introduction}\label{sec1}

Real-world dynamical systems require control methods that ensure safety while executing increasingly complex tasks. Signal Temporal Logic (STL) \cite{stl_ref} provides a formal language for specifying such tasks along with quantitative robustness measures.
A common approach is to encode STL specifications as mixed-integer linear constraints and solve them via MILP \cite{STL_MPC, STL_MILP_bacspinar2019mission, STL_MILP}; related methods using Bézier-curve constraints achieve similar results \cite{STL_verhagen2024temporally}. While effective, these methods suffer from high computational cost and limited scalability. Other approaches \cite{farahani2015robust} successfully incorporated model predictive control with a smaller optimization horizon to solve STL tasks. However, they do not scale well in continuous optimization problems. Recently, reinforcement-learning-based methods \cite{q_rl_stl, rlstl, n_stl} embed STL robustness into reward functions, avoiding explicit optimization; however, they still face scalability issues and lack formal safety guarantees.\\

Control barrier functions (CBFs) are widely used to enforce certified safety \cite{cts_cbf, discrete_CBF}. The system's safety is established by enforcing the invariance of a safe set, a zero superlevel set of CBF, and by designing a controller to keep the system trajectory inside the safe set. Since STL contains temporal constraints and standard CBFs are time-invariant, they cannot directly capture time-varying STL predicates. Time-varying CBFs have been proposed for STL satisfaction \cite{CBF_STL_lindemann2018, STL_CBF_2024cbf}, including extensions to multi-agent systems \cite{lindemann2020barrier}. These approaches typically require hand-crafted CBF templates for each STL interval and rely on quadratic programs that may become infeasible under strict input constraints. For instance, the system starting from a state outside an unsafe set would not be able to satisfy the STL specification within the specified time due to the given input constraints. But in the existing literature, either the states are presumed to be safe initial states simply because they are not explicitly in the unsafe region, or a very small initial state set is considered. In either case, solving the hand-crafted CBF template using a quadratic program would be infeasible under input constraints. Additionally, these works focus on control-affine systems and on fragments of STL specifications that do not consider disjunctions. Related work in \cite{cbf_mpc_, TV_CBF_Receding_horizon_2023} addresses STL satisfaction for linear systems under input constraints, either by encoding STL into CBFs and computing least-violating controls \cite{cbf_mpc_} or by online parameterization of time-varying CBFs \cite{TV_CBF_Receding_horizon_2023}. While promising, these methods are limited to linear system dynamics. In contrast, we develop a framework applicable to any general continuous-time system (including non-affine systems), thereby expanding the class of systems for which STL specifications can be enforced under bounded inputs. Recent developments in neural network-based and data-driven CBF \cite{anand2023formally, Data_CBF_Zamani} have eliminated the need for handcrafted CBF templates, but they focus on time-invariant settings and can not be trivially extended to handle temporal STL predicates. To the best of our knowledge, this is the first work to address the problem of controller synthesis for a class of STL tasks without relying on predefined CBF templates and refining the time-varying set that satisfies the STL task while ensuring input constraints.\\
\\
This letter considers continuous-time systems with a fragment of STL tasks subjected to input constraints. We propose a control strategy that ensures STL satisfaction under input constraints. We iteratively construct time-varying sets that encode the STL semantics, keeping the input constraints into consideration, and also formulate control barrier conditions over those sets. 
We leverage the approximation capability of neural networks to develop a neural network-based time-varying control barrier function and an associated neural network-based controller for the given STL specifications under input constraints. Since neural networks are trained on finite sample datasets, we also provide formal guarantees over the entire state space by proposing a validity condition that ensures STL satisfaction. We validate the effectiveness of the proposed framework by applying it to various continuous-time systems, each with different STL specifications and control limits. Our main contributions are as follows:
\begin{itemize}
    \item We address the controller design problem for a fragment of STL specifications under input constraints.
    \item We propose a unified framework that (i) iteratively constructs time-varying sets encoding STL specifications, (ii) formulates a time-varying control barrier function (TVCBF) over the constructed sets.
    \item We develop a neural network-based TVCBF along with a neural network controller to enforce STL satisfaction. We also derive a validity condition that ensures STL satisfaction over the entire state space, despite training the neural networks on a finite dataset.
    \item Finally, we demonstrate the effectiveness of our proposed work on multiple continuous-time systems (including affine and non-affine systems) subjected to STL specification and input constraints. 
\end{itemize}

This paper is organized as follows: Section \ref{preliminary} presents the class of systems we focus on, introduces signal temporal logic specification, and presents the problem we address in this paper. Section \ref{tvcbf} presents the detailed formulation of time-varying CBF (TVCBF) for a continuous state space and then introduces the neural network-based TVCBF for a finite samples from the state-time space. It also provides theoretical guarantees for STL satisfaction using the neural TVCBF designed over the sampled points to extend it to the continuous state space. In Section \ref{training}, we describe the neural network architecture and the training algorithm for neural networks (both TVCBF and controller networks) and then provide the strategy of iteratively constructing and refining a continuously differentiable time-varying set that encodes the STL specifications. Finally, in Section \ref{results}, we provide simulation results for various continuous-time systems. Section \ref{conclusion} concludes the paper with a summary.

\section{Preliminaries and Problem Formulation}\label{preliminary}
\textbf{Notations:} The sets of real and non-negative real numbers are denoted by $\mathbb{R}$ and $\mathbb{R}_{\geq 0}$, respectively. The set of natural numbers between  $1$ to $N$ is denoted by $[1;N]$. An $n$-dimensional vector space is $\mathbb{R}^n$ and a column vector is $x=[x_1, ...,x_n]^\top\in \mathbb{R}^n$. The symbol $\preceq$ denotes element-wise inequality of vectors. We denote a set of real matrices with $n$ rows and $m$ columns by $\mathbb{R}^{n\times m}$. A continuous function $\alpha:(-a,b)\rightarrow \mathbb{R}$ for $a,b>0$ is called an extended class $\mathcal{K}$ function if it is strictly increasing, $\alpha(0)=0$ and it is denoted as $\mathcal{K}_e$. The notation for partial differentiation of a function $f:\mathbb{R}^n\times \mathbb{R}\rightarrow \mathbb{R}$ with respect to the variables $x\in \mathbb{R}^n$ and $t\in \mathbb{R}$ is $\frac{\partial f}{\partial x}$ and $\frac{\partial f}{\partial t}$, respectively. The $p$-norm is represented using $||\cdot||_p$. An indicator function $\mathbbold{1}_{x\in X}=1$, if $x\in X$, and 0 otherwise. A Lipschitz continuous function $f$ has a Lipschitz constant $L\in\mathbb{R}_{\geq 0}$ if $||f(x_1)-f(x_2)||_2\leq L||x_1-x_2||_2$. 
\subsection{System Description}
Consider a continuous-time nonlinear control system 
\begin{align}
    \Sigma: \dot{\mathbf{x}}(t) = f(\mathbf{x}(t),\mathbf{u}(t)), 
    \label{dyn_eq}
\end{align}
where $\mathbf{x}(t)\in X$, $\mathbf{u}(t)\in U$ are the state and input of the system at time $t\in \mathbb{R}_{\geq 0}$, and $X\subset \mathbb{R}^{n}$, $U\subset \mathbb{R}^m$ are assumed to be compact sets representing state and input constraints. The function $f:X\times U\rightarrow \mathbb{R}^n$ is known and assumed to be a Lipschitz continuous function with respect to $x,u$ over $X, U$, with Lipschitz constants $L_x, L_u$, respectively. Let us define a state trajectory starting from $x_0$ with the input signal $\mathbf{u}$ as $\mathbf{x}_{x_0,\mathbf{u}}$.
\subsection{Signal Temporal Logic (STL)}
Signal Temporal Logic (STL) provides a formal language for specifying spatial, temporal, and logical properties \cite{stl_ref}. An STL formula is composed of predicates combined with temporal and logical operators. Let $\mathbf{x}:\mathbb{R}_{\geq 0} \rightarrow X \subseteq \mathbb{R}^{n}$ be a time-varying signal, and a predicate function $h:X\rightarrow \mathbb{R}$. A predicate is $\mu = \mathsf{true}$ if $h(x(t))\geq 0$, and $\mathsf{false}$, otherwise.
The basic STL formulas are as follows:
\begin{align*}
    \hspace{-0.1cm}\phi ::= \mathsf{true} | \mu | \lnot\phi|\phi_1\wedge\phi_2 | \phi_1\vee\phi_2 | \Box_{[a,b]}\phi | \lozenge_{[a,b]}\phi | \phi_1\mathcal{U}_{[a,b]}\phi_2,  
\end{align*}
where $\mu$ is a predicate, the operators $\lnot$, $\wedge$, and $\vee$ represent the logical negation, conjunction, and disjunction operators, respectively. The temporal operators $\Box$, $\lozenge$, and $\mathcal{U}$ mean `Always', `Eventually', and `Until' operators. The set $[a, b]\subset\mathbb{R}_{\geq 0}$ is the time interval in which the temporal operators are active. The formal semantics of can be found in \cite{stl_ref}. The degree to which a signal $\mathbf{x}$ satisfies an STL specification $\phi$ at time $t$ is quantified by the robustness measure, denoted as $\rho^\phi(\mathbf{x},t) \in \mathbb{R}$. For the STL formulae, the robustness semantics are defined as:\\ 
\begin{align*}
    \rho^\mu(\mathbf{x},t) &= h(\mathbf{x}(t)),\\
    \rho^{\lnot\phi}(\mathbf{x},t) &= -\rho^\phi(\mathbf{x},t),\\
    \rho^{\phi_1\wedge \phi_2}(\mathbf{x},t) &= \min \big(\hspace{-.1em}\rho^{\phi_1}\hspace{-.1em}(\mathbf{x},t),\rho^{\phi_2}\hspace{-.1em}(\mathbf{x},t) \big),\\
    \rho^{\phi_1\vee \phi_2}\hspace{-.1em}(\mathbf{x}, t)  &= \max \big(\hspace{-.1em}\rho^{\phi_1}\hspace{-.1em}(\mathbf{x},t), \rho^{\phi_2}\hspace{-.1em}(\mathbf{x},t) \big),\\
    \rho^{\square_{[a,b]}\varphi}(\mathbf{x},t) &=\underset{t'\in[t+a,t+b]}{\min} \rho^\varphi(\mathbf{x},t'), \\
    \rho^{\lozenge_{[a,b]}\varphi}(\mathbf{x},t) &=\underset{t'\in[t+a, t+b]}{\max} \rho^\varphi(\mathbf{x},t').\\
\end{align*}

We say a signal $\mathbf{x}$ satisfies the STL specification, denoted by $(\mathbf{x},0) \models \phi$ iff $\rho^\phi(\mathbf{x},0)\geq 0$. For brevity, we denote $\rho^\phi(\mathbf{x})\geq 0\equiv\rho^\phi(\mathbf{x},0)\geq 0$ and $\mathbf{x} \models \phi\equiv (\mathbf{x},0) \models \phi$ throughout the paper. For a control system in \eqref{dyn_eq}, an STL specification $\phi$ is satisfiable from the initial state $x_0\in X$ if there exists an input signal $\mathbf{u}$ such that $\rho^\phi(\mathbf{x}_{x_0,\mathbf{u}})\geq 0$.

\subsection{Problem Formulation}
In this paper, we consider the following STL fragment:
\begin{subequations}
    \begin{align}
    \varphi &::= \mathsf{true} | \mu | \lnot \varphi | \varphi_1\wedge\varphi_2 | \varphi_1\vee\varphi_2,\label{STL_formulae_1}\\
    \phi &::= \Box_{[a,b]}\; \varphi | \lozenge_{[a,b]}\;\varphi, \label{STL_formulae_2}\\
    \Phi &::= \bigwedge\limits_{i=1}^{N} \phi_i, \label{STL_formulae}
\end{align}
\end{subequations}
where $\phi_i, i\in\{1,...,N\}$ are STL formulas defined for the interval $[a_i, b_i]$ is of the form $\phi$. The final STL syntax is of the form $\Phi$, such that $\cup_{i}^{N}[a_i, b_i] \subseteq [0,T]$, where $T$ is the total time duration covered by the STL specification $\Phi$.

For the STL formulae mentioned in \eqref{STL_formulae}, the robustness semantics can be defined as $\rho^{\Phi}(\mathbf{x})=\min_{i\in\{1,...,N\}}\rho^{\phi_i}(\mathbf{x})$.

\begin{problem}\label{pr1}
Given an STL specification $\Phi$ of the form \eqref{STL_formulae} for a time duration of $[0,T]$ for a continuous-time nonlinear control system $\Sigma$ in \eqref{dyn_eq}, our objective is to synthesize a controller $\mathbf{u}(t)=\textsl{g}(\mathbf{x}(t),t)$ (if it exists) that ensures the system trajectory $\mathbf{x}_{x_0, \mathbf{u}} $ starting at $x_0$ satisfies the specification $\Phi$ under input constraints $U\subset \mathbb{R}^m$. 
\end{problem}
This letter proposes a framework that iteratively constructs time-varying sets encoding the given STL specification and co-designs a time-varying control barrier function and controller to keep the system within these sets under input constraints, thereby ensuring STL satisfaction. To eliminate the need to predefine the barrier and controller templates, we employ a neural network (NN) approach.


\section{Time-varying Control Barrier Function}\label{tvcbf}
This work designs a controller using control barrier functions to satisfy the STL task of the form \eqref{STL_formulae}. Because the STL specification imposes both temporal and spatial constraints, time-varying control barrier functions (TVCBF) are required. This section reviews TVCBFs, which guarantee STL satisfaction for continuous-time systems. 
\subsection{Time-Varying Control Barrier Function}
The time-varying control barrier function-based approach is used to synthesize a controller that ensures that the system trajectory stays inside a time-varying set for all time. We define an augmented set $W=X\times [0,T]$, where $X\subseteq\mathbb R^n$ and time interval $[0,T], T\in \mathbb{R}_{\geq 0}$. 
\begin{definition}\label{valid_cbf}
    Time-varying Control Barrier Function (TVCBF): A continuously differentiable time-varying function $\mathcal{B}:W\rightarrow \mathbb{R}$ is a control barrier function for a control system $\Sigma$ in \eqref{dyn_eq}, if for a time-varying set $\mathcal{C}(t)\subset W$, there exists a continuous function $\textsl{g}:W\rightarrow U$ such that
\begin{subequations}
    \begin{align}
        &\forall (x,t)\in \mathcal{C}(t), \mathcal{B}(x,t)\geq 0, \label{cbf_1}\\
        &\forall (x,t)\in W\setminus \mathcal{C}(t), \mathcal{B}(x,t)<0, \label{cbf_2}\\
        &\forall (x,t)\in W, \frac{\partial\mathcal{B}}{\partial x} f(x,\textsl{g}(x,t))+\frac{\partial\mathcal{B}}{\partial t} \geq -\alpha(\mathcal{B}(x,t)),\label{cbf_3}
    \end{align}
\end{subequations}
for some class $\mathcal{K}_e$ function $\alpha$. 
\end{definition}

\begin{theorem} \label{th:1}
    For a continuous-time control system $\Sigma$ in \eqref{dyn_eq} and a time-varying set $\mathcal{C}(t)$, suppose there exist a continuously differentiable function $\mathcal{B}:W\rightarrow \mathbb{R}$ and a controller $\textsl{g}:W\rightarrow U$ satisfying conditions \eqref{cbf_1}-\eqref{cbf_3}. Then, the system trajectory $\mathbf{x}_{x_0,\mathbf{u}}$ starting from $(x_0,0)\in \mathcal{C}(0)$ with $\mathbf{u}(t)=\textsl{g}(\mathbf{x}(t),t)$, will always stay in $\mathcal{C}(t)$, i.e., $\mathbf{x}_{x_0,\mathbf{u}}(t)\in \mathcal{C}(t), \forall t\in[0,T]$.
\end{theorem}
\begin{proof}
    Assuming that the system starts at $(x_0,0)\in \mathcal{C}(0)$ implies $\mathcal{B}(x_0,0) \geq 0$ (as per \eqref{cbf_1}). Now, considering condition \eqref{cbf_3}, there exists a control signal $\textsl{g}(\mathbf{x}(t),t)$ such that $\dot{\mathcal{B}}(\mathbf{x}(t),t)\geq -\alpha(\mathcal{B}(\mathbf{x}(t),t))$, where $\alpha(\cdot)$ is a class $\mathcal{K}_e$ function. Suppose $b(t)=\mathcal{B}(\mathbf{x}(t),t)$, then $\dot{b}(t)\geq -\alpha(b(t))$. Let $\beta$ be the solution of the equation to $\dot{\beta}(t)=-\alpha(\beta(t))$, and $\beta(0)=b(0)$. Using Comparison lemma \cite[Chapter 3]{khalil2002nonlinear}, $b(t)\geq \beta(t), \forall t\in [0,T]$. Since $b(0)\geq 0$, and $\beta(t)$ is non-negative for all $t$, we have $b(t) = \mathcal{B}(\mathbf{x}(t),t)\geq \beta(t) \geq 0, \forall t\in [0,T]$. Therefore, we conclude that $(\mathbf{x}(t),t) \in \mathcal{C}(t), \forall t\in [0,T]$ (By Definition \ref{valid_cbf}). 
\end{proof}
\subsection{TVCBF for STL Specifications}
Let us consider the STL specification $\Phi$ of the form \eqref{STL_formulae} be defined over the time interval $[0,T]$, and each $\phi_i, i\in \{1, ..., N\}$, is an eventually or always operator with the corresponding time interval $[a_i, b_i]$. Let us denote the set of STL sub-formulae with the eventually operator as $\Phi_\lozenge = \{\phi_i\mid\phi_i=\lozenge_{[a_i,b_i]}\varphi_i\}$, and the sub-formulae with the always operator as $\Phi_{\square}=\{\phi_i\mid\phi_i=\square_{[a_i,b_i]}\varphi_i\}$, 
such that $\cup_{i=1}^N I_i\subseteq [0,T]$, where
\begin{align}
I_i=\begin{cases}
[a_i, b_i], &\text{ if } \phi_i\in \Phi_\square,\\
[t^*, t^*+\delta]\subset [a_i,b_i], &\text{ if } \phi_i\in \Phi_\lozenge.\\
\end{cases}
\end{align}
{The variables $t^*, \delta$ are selected such that:} $a_i\leq t^*< t^*+\delta\leq b_i$, and $\delta>0$. 
The set of all active predicate components of the STL formula at time $t$ is given by $\Phi_{a}(t) = \{\varphi_i\mid t\in I_i\}$.
\begin{assumption}\label{STL_assumption}
    We assume that at least one predicate is active at any given time, i.e.,$ \Phi_a(t)\neq \emptyset, \forall t\in[0,T]$, {on account of the state space constraints imposed on the system}.
\end{assumption}
We set one of the STL sub-formulae describing state space constraints for the state $x\in X$ in the entire time duration $[0,T]$, without compromising on the given STL task, resulting in an STL formula of the form:
    \begin{align}
        \Phi &=\square_{[0,T]}(D||x-x_c||_p\leq 1)\land\Phi_1,
    \end{align}
    where $x_c$ is the center of the state space constraint $A\subset X$, $||\cdot||_p$ denotes $p$-norm with $p=1,2,\infty$, the constant $D=1/r$, for a circular state space constraint of radius $r$, $D = \text{diag}(1/w_i)$ is a diagonal matrix for a rectangular state space bound with $w_i$ as half the width of the constraint along each state dimension of $A\subset X$, and $\Phi_1$ is of the form \eqref{STL_formulae}.
    
We now take the augmented set for finite-time $W = X\times[0,T]$ and define the time-dependent set of states that satisfy the STL specification as follows:
\begin{align}
    \hspace{-0.2cm}\mathcal{S}^\Phi(t) = \{(x,t)\in W&| \min_{\varphi_i}\big(\rho^{\varphi_i}(x)\big)  \geq 0,\forall{\varphi_i\in\Phi_a(t)}\},
    \label{safe_set_def}
\end{align}
where $\varphi_i$ is the non-temporal predicate of the form \eqref{STL_formulae_1}.

\begin{theorem}\label{th:2}
For a continuous-time control system as in \eqref{dyn_eq} and an STL specification $\Phi$ of the form \eqref{STL_formulae} satisfying Assumption \ref{STL_assumption}, suppose that there exist a continuously differentiable function $\mathcal{B}$ and a controller $\textsl{g}:X\times [0,T]\rightarrow U$ satisfying conditions \eqref{cbf_1}-\eqref{cbf_3} in Definition \ref{valid_cbf}, with some set $\mathcal{C}(t)=\mathcal{C}^\Phi(t)\subset \mathcal{S}^\Phi(t)$, where $\mathcal{S}^\Phi(t)$ is defined as \eqref{safe_set_def} for STL specification $\Phi$. Then, the system trajectory $\mathbf{x}_{x_0,\mathbf{u}}$ starting from $(x_0,0)\in \mathcal{C}^\Phi(0)$ with $\mathbf{u}(t)=\textsl{g}(\mathbf{x}(t),t)$ will always stay in $\mathcal{C}^\Phi(t)$, i.e., $\mathbf{x}_{x_0,\mathbf{u}}(t)\in\mathcal{C}^\Phi(t),\forall t\in [0,T]$, and the STL specification $\Phi$ is satisfied by $\mathbf{x}_{x_0,\mathbf{u}}$, i.e., $\mathbf{x}_{x_0,\mathbf{u}} \models\Phi$.
\end{theorem}
\begin{proof}
By Theorem \ref{th:1}, the system trajectory starting from $(x_0,0)\in \mathcal{C}(0)$ under controller $\textsl{g}$ will always stay in $\mathcal{C}^\Phi(t)$, i.e., $\mathbf{x}_{x_0,\mathbf{u}}(t)\in\mathcal{C}^\Phi(t),\forall t\in [0,T]$. By the construction of $\mathcal{C}^\Phi(t)\subset \mathcal{S}^\Phi(t)$ in \eqref{safe_set_def}, we have $\min_{\varphi_i}\big(\rho^{\varphi_i}(\mathbf{x}(t))\big) \geq 0,\forall \varphi_i\in\Phi_a(t), \forall t\in [0,T]$. This implies that the robustness corresponding to active STL fragments at each time is positive; consequently, $\mathbf{x}_{x_0,\mathbf{u}}\models \Phi$.
\end{proof}
\subsection{Neural Network-based Time-Varying CBF (N-TVCBF)} \label{section_N_TVCBF}
In this section, we synthesize TVCBF along with the controller for a system to satisfy an STL specification. 
The set $\mathcal{S}^\Phi(t)$ formed using \eqref{safe_set_def} is not continuously differentiable, and therefore we still have a major challenge to construct a set $\mathcal{C}^\Phi(t)\subset\mathcal{S}^\Phi(t)$ in which the barrier function $\mathcal{B}(x,t)$ is continuously differentiable. Additionally, due to input constraints, the trajectory $\mathbf{x}_{x_0,u}(t)$, starting at some $x_0\in\mathcal{S}^\Phi(0)$ might leave the STL safe set $\mathcal{S}^\Phi(t)$ at some time $t\in [0,T]$ and therefore not satisfy the specification $\Phi$. We propose an iterative refinement approach in the next section to construct a set that ensures STL satisfaction, along with feasibility with respect to input constraints and continuity in time. In this section, we focus on TVCBF assuming that the appropriate time-varying set $\mathcal{C}^\Phi(t)$ is available.
\begin{lemma}
    The continuous-time system $\Sigma$ in \eqref{dyn_eq} with trajectory starting from $(x_0,0)\in \mathcal{C}^\Phi(0)$ satisfies the STL specification $\Phi$ if the following conditions hold with $\eta\leq 0$.
\begin{align}
    &\max(q_k(x,t))\leq \eta, k\in[1;3], \forall (x,t)\in W,\label{rop}
\end{align}
where
\begin{align}
    q_1(x,t) &= -\mathcal{B}(x,t)\mathbbold{1}_{(x,t)\in \mathcal{C}^\Phi(t)},\nonumber\\
    q_2(x,t) &= (\mathcal{B}(x,t)+\lambda)\mathbbold{1}_{(x,t)\in W\setminus \mathcal{C}^\Phi(t)},\nonumber\\
    q_3(x,t) &= -\frac{\partial\mathcal{B}}{\partial x}f(x,\textsl{g}(x,t)) - \frac{\partial\mathcal{B}}{\partial t} -\alpha(\mathcal{B}(x,t)), \label{q_k}
\end{align}
and $\lambda>0$ enforces strict inequality in \eqref{cbf_2}.
\end{lemma}
\begin{proof}
    The first inequality for $q_1$ with $\eta\leq 0$ ensures that the CBF $\mathcal{B}(x,t)\geq 0$ in $(x,t)\in\mathcal{C}^\Phi(t)$. Additionally, $q_2$ with $\eta\leq 0$ ensures a strict negative value of CBF in $(x,t)\in W\setminus\mathcal{C}^\Phi(t)$. The inequality $q_3$ with $\eta\leq 0$ ensures the confinement of the system trajectory within the set $\mathcal{C}^\Phi(t)$. Using Theorem \ref{th:1} and \ref{th:2}, all these inequalities together ensure that a system trajectory starting from $(x_0,0)\in \mathcal{C}^\Phi(0)$ with $\mathbf{u}(t)=\textsl{g}(\mathbf{x}(t),t)$ satisfies the STL specification $\Phi$.
\end{proof}
One major challenge in solving these inequalities is the infinite number of constraints resulting from the continuous state-space. To overcome this, we use a finite number of samples from the set $W$. 
We generate a set of $N$ samples $s^{(r)} = (x,t)^{(r)}$, where $r\in [1;N]$, such that 
\begin{align}
   \forall(x,t)\in W, \exists s^{(r)}\in W,||(x,t)-s^{(r)}||_2\leq \epsilon. \label{eps_x_t}
\end{align} Now, we consider the set $\tilde{\mathcal{S}}^\Phi(t)=\{s^{(r)}\mid s^{(r)}\in \mathcal{S}^\Phi(t), r\in[1;N]\}$. A smaller value of $\epsilon$ ensures dense sampling such that $\tilde{\mathcal{S}}^\Phi(t)\cap\mathcal{S}^\Phi(t)\neq \emptyset$.
We assume that we have the set $\tilde{\mathcal{C}}^\Phi(t^{(r)})\subset \tilde{\mathcal{S}}^\Phi(t^{(r)})$ which is a zero-superlevel set of $\mathcal{B}((x,t)^{(r)})$. In addition, instead of pre-defining the TVCBF and controller template, we approximate them with NNs and denote them as $\mathcal{B}_{\theta_1}$ and $\textsl{g}_{\theta_2}$, both parameterized by the trainable parameters $\theta_1$ and $\theta_2$, respectively. 

\begin{assumption} \label{assumption}
    The candidate N-TVCBF $\mathcal{B}_{\theta_1}$ and its derivative are assumed to be Lipschitz continuous with Lipschitz constants $L_b$ and $L_{db}$, respectively \cite{basu2025lemma}. The controller NN $\textsl{g}_{\theta_2}$ has the Lipschitz constant $L_\textsl{g}$. Additionally, the 2-norm of partial derivatives $||\frac{\partial\mathcal{B}}{\partial x}, \frac{\partial\mathcal{B}}{\partial t}||_2$, function $f(x,u)$ are bounded by $M_{b}, M_f$ respectively, i.e., $\sup_{x,t}||(\frac{\partial\mathcal{B}}{\partial x},\frac{\partial\mathcal{B}}{\partial t})||_2\leq M_{b}, \sup_{(x,u)}||f(x,u)||_2\leq M_f $.
\end{assumption}
\begin{lemma} \cite[Ex. 3.3]{khalil2002nonlinear}
    If two functions $h_1$ and $h_2$ are Lipschitz continuous with constants $L_1$ and $L_2$, respectively, and are bounded by $\sup ||h_1||_2\leq M_1$ and $\sup||h_2||_2\leq M_2$, then their product $h_1h_2$ is also Lipschitz continuous with Lipschitz constant $M_1L_2+M_2L_1$. \label{lemma}
\end{lemma}

\begin{theorem}\label{th:3}
Consider a continuous-time control system \eqref{dyn_eq} with compact state and input sets $X$ and $U$, and an STL specification $\Phi$ of the form \eqref{STL_formulae} satisfying Assumption \ref{STL_assumption}. With Assumption \ref{assumption}, the system trajectory $\mathbf{x}_{x_0,\mathbf{u}}$ starting at $x_0$ and $\mathbf{u}(t)=\textsl{g}_{\theta_2}(\mathbf{x}(t),t)$, is said to satisfy $\Phi$ under the time-varying barrier $\mathcal{B}_{\theta_1}$, controller $\textsl{g}_{\theta_2}$ trained over the sampled points as in \eqref{eps_x_t} if the following holds with $\eta+\mathcal{L}\epsilon\leq 0$:
\begin{align}
    \hspace{-0.2cm}\max (q_k(s^{(r)}))\leq \eta, k\in[1;3], \forall s^{(r)}\in W, \forall r\in[1;N],\label{ineq}
\end{align} 
where $\epsilon$ is as defined in \eqref{eps_x_t}, $q_1,q_2,q_3$ are as defined in \eqref{q_k}. The maximum of the Lipschitz constants of $q_k, k\in[1;3]$ in \eqref{q_k} is $\mathcal{L}= \max\{L_1, L_2, L_3\}$, where $L_1=L_2=L_b, L_3=L_{db}(M_f+1)+M_b(L_x+L_uL_\textsl{g})+\alpha L_b$, and the class $\mathcal{K}_e$ function is assumed to be of the form $\alpha(z)=\alpha z, \alpha> 0$.
\end{theorem}
\begin{proof}
    First, we show that, under condition $\eta+\mathcal{L}\epsilon\leq 0$, the constructed $\mathcal{B}_{\theta_1}$ via solving the inequalities in \eqref{ineq} satisfy \eqref{cbf_1}-\eqref{cbf_3} for the entire state space $ X$ and time space $[0,T]$.
    Using \eqref{eps_x_t}, Assumption \ref{assumption}, and Lemma \ref{lemma}, we obtain:

    \begin{align*}(i) \quad&
    \forall (\mathbf{x}(t),t)\in\mathcal{C}^\Phi(t), \exists s^{(r)}\in \tilde{\mathcal{C}}^\Phi(t^{(r)}), {r\in[1;N]},  {s.t., ||(\mathbf{x}(t),t)-s^{(r)}||\leq \epsilon}\\
        &q_1(\mathbf{x}(t),t) = q_1(\mathbf{x}(t),t)-q_1(s^{(r)})+q_1(s^{(r)})\\
        &\quad\quad\quad = (-\mathcal{B}(\mathbf{x}(t),t)+\mathcal{B}(s^{(r)})) - \mathcal{B}(s^{(r)})\\
        &\quad\quad\quad \leq L_b\hspace{0.1cm} \epsilon+\eta^* \leq \mathcal{L}\epsilon+\eta^* \leq 0. \\
    (ii)\quad&
    \forall (\mathbf{x}(t),t)\in W\setminus\mathcal{C}^\Phi(t), \exists s^{(r)}\in W\setminus \tilde{\mathcal{C}}^\Phi(t^{(r)}), r\in[1;N], {s.t., ||(\mathbf{x}(t),t)-s^{(r)}||\leq\epsilon}\\
        &q_2(\mathbf{x}(t),t) = q_2(\mathbf{x}(t),t)-q_2(s^{(r)})+ q_2(s^{(r)}) \\
        &\quad\quad\quad\quad \leq L_b\hspace{0.1cm}\epsilon +\eta^*\leq \mathcal{L}\epsilon+\eta^*\leq 0.
    \\
    (iii)\quad
    &\forall (\mathbf{x}(t),t)\in W, \exists s^{(r)}\in W, r\in[1;N], {s.t., ||(\mathbf{x}(t),t)-s^{(r)}||\leq\epsilon}\\
        &q_3(\mathbf{x}(t),t) = q_3(\mathbf{x}(t),t) - q_3(s^{(r)}) + q_3(s^{(r)})\\
        &\quad\quad\quad\quad\hspace{0.1cm} =-\frac{\partial\mathcal{B}}{\partial \mathbf{x}}f(\mathbf{x}(t),\textsl{g}(\mathbf{x}(t),t)) -\frac{\partial\mathcal{B}}{\partial t}- \alpha \mathcal{B}(\mathbf{x}(t),t) \\&\quad\quad\quad\quad\quad \quad+\frac{\partial\mathcal{B}}{\partial x^{(r)}}f(x^{(r)},\textsl{g}(s^{(r)})) + \frac{\partial\mathcal{B}}{\partial t^{(r)}} + \alpha \mathcal{B}(s^{(r)}) \\
        & \quad\quad\quad\quad\quad\quad-\frac{\partial\mathcal{B}}{\partial x^{(r)}}f(x^{(r)},\textsl{g}(s^{(r)})) -\frac{\partial\mathcal{B}}{\partial t^{(r)}}- \alpha \mathcal{B}(s^{(r)}) \\
        &\quad\quad\quad\quad \hspace{0.1cm}\leq M_fL_{db}\epsilon + M_{b}(L_x+L_uL_\textsl{g})\epsilon +L_{db}\epsilon + \alpha L_b\epsilon+\eta^*\\
        &\quad\quad\quad\quad \hspace{0.1cm} \leq \big(L_{db}(M_f+1)+M_b(L_x+L_uL_\textsl{g})+\alpha L_b\big)\epsilon+\eta^*\\
        &\quad\quad\quad\quad \hspace{0.1cm}\leq \mathcal{L}\epsilon+\eta^* \leq 0.
    \end{align*}
    This implies that if condition \eqref{ineq} is satisfied with $\mathcal{L}\epsilon+\eta\leq 0$, so are the conditions in continuous space in \eqref{rop}. Therefore, the N-TVCBF $\mathcal{B}_{\theta_1}$ satisfies \eqref{cbf_1}-\eqref{cbf_3} and by Theorem \ref{th:2}, the controller ensures STL specification $\Phi$ is satisfied.
\end{proof}

 
\section{Training of N-TVCBF and Controller}\label{training}
This section presents the neural network architecture, the construction of set $\tilde{\mathcal{C}}^\Phi(t)$, the loss functions designed for the TVCBF constraints \eqref{cbf_1}-\eqref{cbf_3}, and the training process used to ensure formal guarantees.
\subsection{Neural Network Architecture}
\begin{figure}
    \centering
    \includegraphics[width=0.5\linewidth]{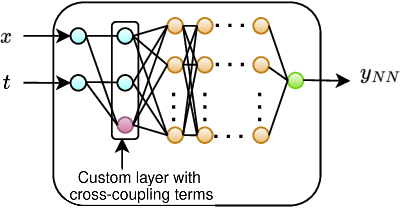}
    \caption{Neural network architecture}
    \label{fig:NN_arch}
\end{figure}

We denote the neural network (NN) architecture as $\{n^0, n^c,\{n^l\}^l,n_o\}$, consisting of an input layer with $n^0$ neurons, a custom layer with $n^c$ neurons, $l$ hidden layers of width $n^l$, and an output layer of size $n_o$. The custom layer introduces explicit cross-coupling between the inputs $t$ and $x=[x_1, x_2,..., x_n]^\top\in {X}$, e.g., $\{tx_1, tx_2,..., tx_n\}$, or $\{e^{a_1t}x_1,e^{a_2t}x_2,..., e^{a_nt}x_n\}$ (cf. Figure \ref{fig:NN_arch}), which facilitates the TVCBF approximation during training. Each layer uses weights $w_i\in\mathbb{R}^{n^{i+1}\times n^i}$, biases $b_i\in\mathbb{R}^{n^{i+1}}$, and a smooth activation $\sigma(\cdot)$ (e.g., Softplus, Tanh, Sigmoid, SiLU) to enable the computation of partial derivatives of the NN with respect to its input ($\frac{\partial\mathcal{B}}{\partial x}, \frac{\partial\mathcal{B}}{\partial t}$). The resulting NN function is obtained by recursively applying the activation function in the hidden layers. The output of each layer in the NN is given as $z_{k+1} = \Sigma_{k}(w_kz_k+b_k), \forall k\in\{0,1,..., l-1\}$, where $\Sigma_i:\mathbb{R}^{n^i}\rightarrow\mathbb{R}^{n^i}$ is $\Sigma_i(z_i) = [\sigma(z_{i}^{1}),...,\sigma(z_{i}^{n^i})]$ with $z_i$ denoting the concatenation of outputs $z_i^j, j\in \{1,2,...,n^i\}$ of the neurons in the $i$-th layer. 
For the N-TVCBF, the output is $y_{NN}(z_l) = w_lz_l+b_l$, whereas for the controller NN, which is designed to satisfy the input constraints $U=\{u\in \mathbb{R}^m\mid \text{lb}\preceq u\preceq \text{ub}\}$, we bound the output between `lb' and `ub' using the HardTanh activation function as $y_{NN}(z_l)=\text{HardTanh}(w_{l}z_{l} + b_{l})_{\text{lb}}^{\text{ub}}$. The output of $\text{HardTanh}(x)_{\text{lb}}^{\text{ub}}$ is $ \text{lb} \text{ if }x<\text{lb}$, or $\text{ub} \text{ if }x>\text{ub}$, or $x$, otherwise. The overall trainable parameter of the NN is $\theta = [w_0, b_0,..., w_l, b_l]$. For an $n$-dimensional system with $m$ inputs, the architectures are: N-TVCBF $B_{\theta_1}: \{n+1, 2n+1, \{n^l\}^l,1\}$ and controller $g_{\theta_2}:\{n+1, 2n+1,\{n^l\}^l,m\}$ with trainable parameters $\theta_1$ and $\theta_2$.
\subsection{Training Algorithm with Formal Guarantees}\label{loss_fun_form}
To solve Problem \ref{pr1}, we jointly train two neural networks to approximate the TVCBF $\mathcal{B}_{\theta_1}$ and controller $\textsl{g}_{\theta_2}$ so that the associated loss functions for constraints \eqref{ineq} converge. This section summarizes the training algorithm.\\
\textbf{INPUTS: }System dynamics $\Sigma$ satisfying Assumption \ref{assumption} and an STL specification satisfying Assumption \ref{STL_assumption}.\\
\textbf{STEP 1: }Generate $N$ samples, build the dataset $\tilde{\mathcal{S}}^\Phi(t)$ using $\Phi$, and initialize $\tilde{\mathcal{C}}_0^\Phi(t)=\tilde{\mathcal{S}}^\Phi(t)$.\\
\textbf{STEP 2: }Select number of training epochs, Lipschitz bounds ($L_b, L_{dB}, L_{\textsl{g}}$), $M_b$, NN hyperparameters ($l, n_l,$ activation function, optimizer, scheduler), termination criteria (convergence of $L_{cbf}$ to zero or maximum epochs). Initialize $i=0, \lambda >0$, $\eta = -L_{max}\epsilon$, the trainable parameters $\theta_1, \theta_2, \Gamma$.\\
\textbf{STEP 3: }Training starts here:\\
\quad\quad \textbf{(i)} Create batches of training data from $\tilde{\mathcal{C}}^\Phi_i(t)$.\\
\quad \quad \textbf{(ii)} Find batch loss $L_{cbf}=k_1L_1+k_2L_2+k_3L_3 $, with $k_1, k_2, k_3>0$ and 
\begin{align}
    &L_1(\theta_1) = \sum_{s^{(r)}\in \tilde{\mathcal{C}}_i^\Phi(t)} ReLU\big(-\mathcal{B}_{\theta_1}(s^{(r)}) - \eta\big),\label{loss1}\\
    &L_2(\theta_1) = \sum_{s^{(r)}\in W\setminus \tilde{\mathcal{C}}_i^\Phi(t)} ReLU\big(\mathcal{B}_{\theta_1}(s^{(r)}) +\lambda- \eta\big),\label{loss2}\\
    &L_3(\theta_1, \theta_2)=\sum_{s^{(r)}\in W} \hspace{-0.2cm}ReLU\bigg(-\frac{\partial \mathcal{B}_{\theta_1}}{\partial x^{(r)}}f\big(x^{(r)}, \textsl{g}_{\theta_2}(s^{(r)})\big)-\frac{\partial \mathcal{B}_{\theta_1}}{\partial t^{(r)}} -\alpha\big(\mathcal{B}_{\theta_1}(s^{(r)})\big) -\eta \bigg),\label{loss3}
\end{align}
where $ReLU(z) = \max(0,z), \alpha(x)=\alpha z, \alpha>0$.\\
\quad \quad \textbf{(iii)} Update $\theta_1^i, \theta_2^i$ using optimizer (ADAM) \cite{adam}.\\
\quad \textbf{STEP 4:} To ensure Assumption \ref{assumption} and train NNs that are Lipschitz bounded, we use the Lemma adopted from \cite[Lemma 4.1]{basu2025lemma} and satisfy the linear matrix inequalities (LMIs) corresponding to the Lipschitz constants of different networks. We formulate the loss function 
\begin{align}
    &L_M(\Theta, \Gamma)=-c_{l_1}\log\det(M_{L_{b}}(\theta_1, \gamma_1))-c_{_{l_2}}\hspace{-0.1cm}\log\det(M_{L_{dB}}(\hat \theta_1, \hat\gamma_1))\hspace{-0.05cm}-\hspace{-0.05cm}c_{_{l_3}}\hspace{-0.1cm}\log\det(M_{L_{\textsl{g}}}(\theta_2, \gamma_2)), \label{lmi_loss}
\end{align}
where $\Theta=[\theta_1, \theta_2]$, $c_{_{l_1}}, c_{_{l_2}}, c_{_{l_3}}>0$ are weights for sub-loss LMIs, $\Gamma = [\gamma_1, \hat\gamma_1,\gamma_2]$ and $M_{L_{b}}(\theta_1, \gamma_1), M_{L_{dB}}(\hat\theta_1, \hat\gamma_1), M_{L_{\textsl{g}}}(\theta_2, \gamma_2)$ are matrices corresponding to the bounds $L_b, L_{dB}$ and $L_\textsl{g}$ respectively, computed as per \cite[Lemma 4.1]{basu2025lemma}. Using the computed Lipschitz constants and $L_{max}$ as given in Theorem \ref{th:3}, we update $\eta=-L_{max}\epsilon$. \\
\textbf{STEP 5: ($\mathsf{STL Safe Set Refinement}$)} At every $n$-th epoch, update the set 
\begin{align}
    \tilde{\mathcal{C}}_i^\Phi(t^{(r)})&= \tilde{\mathcal{C}}_{i-1}^\Phi(t^{(r)})\setminus\{s^{(r)}\in\tilde{\mathcal{C}}_{i-1}^\Phi(t^{(r)})\mid \mathcal{B}_{\theta_1}(s^{(r)})<0 \nonumber\\&\text{ or }||(\frac{\partial \mathcal{B}_{\theta_1}(s^{(r)})}{\partial x^{(r)}}, \frac{\partial \mathcal{B}_{\theta_1}(s^{(r)})}{\partial t^{(r)}})||\geq M_b\}, \label{safesetrefine}
\end{align} 
where $M_b$ is the maximum value of $||(\frac{\partial \mathcal{B}_{\theta_1}(s^{(r)})}{\partial x^{(r)}}, \frac{\partial \mathcal{B}_{\theta_1}(s^{(r)})}{\partial t^{(r)}})||_2$ (Assumption \ref{assumption}). For rest of the epochs, we keep $\tilde{\mathcal{C}}_i^\Phi(t^{(r)})=\tilde{\mathcal{C}}_{i-1}^\Phi(t^{(r)})$. In this step, we let the samples $s^{(r)}\in \tilde{\mathcal{C}}_{i-1}^\Phi(t^{(r)})$ that have either negative barrier value or large gradient be excluded from $\tilde{\mathcal{C}}_{i}^\Phi(t^{(r)})$ in next iteration. This results in a set in which the barrier $\mathcal{B}_{\theta_1}$ is continuously differentiable. In Figure \ref{fig:non_affine}(a)-(c), we plot the initial safe set ($\tilde{\mathcal{C}}_{0}^\Phi$) and the refinement done at 20th and 40th epochs. We observe that the set {$\tilde{\mathcal{C}}_{40}^\Phi(t)\subset \tilde{\mathcal{C}}_{20}^\Phi(t)\subset \tilde{\mathcal{C}}_{0}^\Phi(t)$.} At the end of training, although we have the input constraints, a trajectory starting at the refined safe set $\tilde{\mathcal{C}}_{Epochs}^\Phi(0)$ will be able to remain in the refined safe set $\tilde{\mathcal{C}}^\Phi_{Epochs}(t^{(r)})$.\\
\textbf{STEP 6: } Increment 
$i$ and repeat \textbf{STEPS 3–5} until termination criteria is satisfied. \\
\textbf{STEP 7: } If losses converge to zero (approximately to $10^{-6}$ to $10^{-4}$), return NNs $\mathcal{B}_{\theta_1}, \textsl{g}_{\theta_2}$, else restart from \textbf{STEP 1}.
\begin{remark}
    {The algorithm lacks a general convergence guarantee, but strategies like reducing the discretization parameter `$\epsilon$' (or increasing the number of samples)\cite{zhao2021learning} or adjusting NN hyper-parameters (architecture, learning rate) \cite{nn_lr} can improve convergence of loss.}
\end{remark}
\begin{algorithm}
\caption{\enskip NN Training}
\label{NN_training}
\begin{algorithmic}[1]
    \Require System dynamics $\Sigma$, STL specification: $\Phi$
    \Ensure $\mathcal{B}_{\theta_1}, \textsl{g}_{\theta_2}, \eta$
    \State Create dataset $\tilde{\mathcal{S}}^\Phi(t)$ using $\Phi$
    \State Initialize $\tilde{\mathcal{C}_0}^\Phi(t)=\tilde{\mathcal{S}}^\Phi(t)$
    
    \State Select: Number of training epochs `$Epochs$', NN hyperparameters ($l,n_l$, activation function, optimizer, scheduler), desired Lipschitz bounds $(L_b, L_{dB}, L_\textsl{g}), M_b,k_1, k_2, k_3$, training termination criteria (until loss goes to zero or maximum epochs is attained).
    \State Initialize: {$i=0$}, $\lambda>0, \eta=-L_{max}\epsilon$, Trainable parameters ($\theta_1, \theta_2, \Gamma$)
    \For{$i\leq Epochs$ (Training starts here)}
        \State Create batches of training data from $\tilde{\mathcal{C}}_i^\Phi(t_d)$
        \State Find batch loss $L_{cbf}=k_1L_1+k_2L_2+k_3L_3$ using \eqref{loss1}-\eqref{loss3}
        \State Use optimizer (such as ADAM \cite{adam}) to update 
        $\theta_{1}^i, \theta_{2}^i$
        \State Find the Lipschitz constants of the networks using the ECLipsE tool \cite{eclipse}
        \State Find LMI loss $L_M$ using \eqref{lmi_loss} and optimize $\Gamma$.
        \State Use the computed Lipschitz constants to update $\eta=-L_{max}\epsilon$
        \State $\mathsf{STL Safe Set Refinement:}$ Update the set $\tilde{\mathcal{C}}_i^\Phi(t)$ using \eqref{safesetrefine} 
        
    \EndFor
    \State \textbf{return} $B_{\theta_1}, \textsl{g}_{\theta_2}, \eta $
\end{algorithmic}
\end{algorithm}

\begin{theorem}\label{th:4}
    Consider a continuous-time system \eqref{dyn_eq}, with compact state and input sets $X$ and $U$, and an STL specification $\Phi$ of the form \eqref{STL_formulae}, satisfying Assumption \ref{STL_assumption}, over the time interval $[0,T]$. 
    Let $\mathcal{B}_{\theta_1}(\mathbf{x}(t),t)$ be the trained N-TVCBF with the corresponding controller $\textsl{g}_{\theta_2}(\mathbf{x}(t),t)$, such that the loss $L_{cbf}$ [in Algorithm 1, Step 3, Line 7] is minimized. If the loss $L_{cbf}$ goes to zero and $L_M(\Theta,\Gamma)\leq 0$, then starting at any point in the set $\mathcal{C}^\Phi(0)\subset \mathcal{S}^\Phi(0)$, the trained controller NN $\textsl{g}_{\theta_2}$ ensures that the STL specification is satisfied.
\end{theorem}
\begin{proof}
    The loss $L_{cbf} = 0 $ implies that the solution to the finite inequality conditions is obtained with $\eta=-L_{max}\epsilon < 0$ (as taken in Algorithm 1-line 4 and updated in Line 11). Additionally, the loss $L_M(\Theta, \Gamma)\leq 0$ implies the satisfaction of Assumption \ref{assumption} with the predefined Lipschitz constants. Hence, using Theorem \ref{th:3}, the controller $\textsl{g}_{\theta_2}$ ensures that the STL specification is satisfied when the system is initialized in $(x_0,0)\in \tilde{\mathcal{C}}^\Phi(0)\subset \tilde{\mathcal{S}}^\Phi(0)$.
\end{proof}
\section{Simulation Results}\label{results}
In this section, we validate the proposed method using mecanum, pendulum, spacecraft and a non-affine scalar system simulations for various STL tasks. We use $\mathcal{K}_e$ function $\alpha (x)=\alpha x$, where $\alpha>0$. The NNs have fixed architecture parameters $n^0, l, n^l, n_o$ (input size, number of hidden layers, hidden-layer width, and output size) and a smooth activation function in all hidden-layers. 


\subsection{Non-affine System}
We consider a control non-affine system of the form \eqref{dyn_eq} 
\begin{align}
    \dot{\mathbf{x}}=
    a(\sin \mathbf{x} + \tan\mathbf{u}),\label{non_affine}
\end{align}
where $a = 0.5$, input $\mathbf{u}$ is bounded within $[-0.5,0.5]$. The neural networks $\mathcal{B}_{\theta_1}, \textsl{g}_{\theta_2}$ are trained to satisfy the following STL specification:

\begin{align}
    \Phi_1 =& \square_{[0,15]}(|\mathbf{x}|\leq \pi/3) \land \square_{[6,8]}(-\pi/3\leq \mathbf{x}\leq -\pi/15) \land \square_{[12.5,15]}\pi/15\leq \mathbf{x}\leq \pi/3). \label{phi_1}
\end{align}
\begin{figure*}[h!]
    \includegraphics[width=\linewidth,trim={1.0cm 0 2cm 0.5cm},clip]{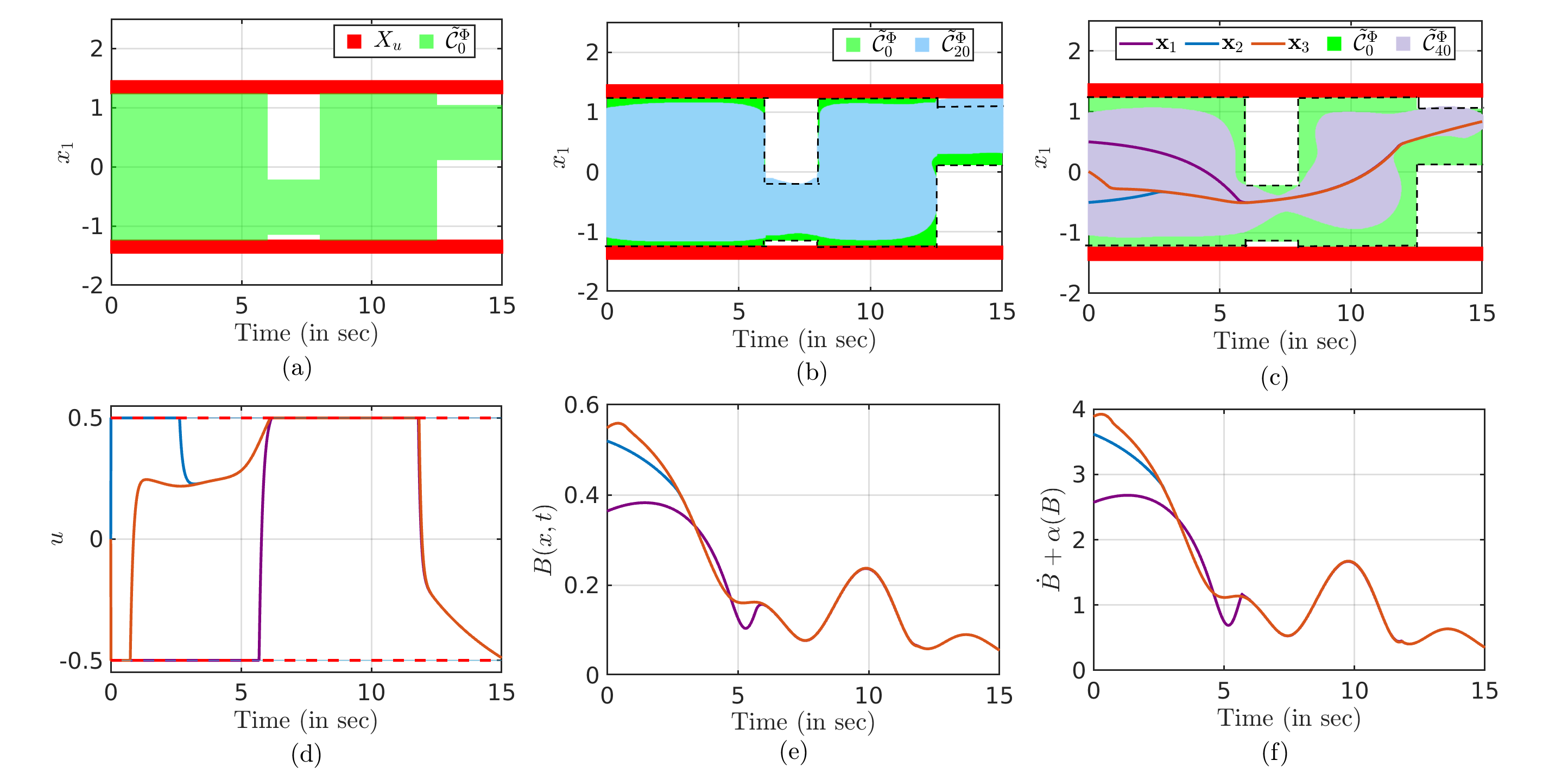}
    \caption{STL Safe set refinement during training at epoch $i=0, 20, 40$ is shown in green ($\tilde{\mathcal{C}}_{0}^\Phi$), light blue ($\tilde{\mathcal{C}}_{20}^\Phi$) and light purple color ($\tilde{\mathcal{C}}_{40}^\Phi$), respectively. The system \eqref{non_affine} satisfying $\Phi_1$ \eqref{phi_1} with state trajectory starting at $x_1(0)=0.5$ (purple), $x_2(0)=-0.5$ (blue), $x_3(0)=0.1$ (brown), N-TVCBF $\mathcal{B}_{\theta_1}(\mathbf{x}(t),t)$ and controller $\textsl{g}_{\theta_2}(\mathbf{x}(t),t)$ satisfies \eqref{cbf_1}-\eqref{cbf_3}, keeping control input within limits (dotted red lines)}
    \label{fig:non_affine}
\end{figure*}
The state space is $X=[-\pi/2, \pi/2]$. The discretization parameter for data sampling is $\epsilon=0.001$. The set $\tilde{\mathcal{S}}^\Phi(t)$ is generated by the predicates formed using the STL $\Phi_1$ (Refer to Section \ref{section_N_TVCBF} and \ref{loss_fun_form}). For example, $\forall t\in [6,8]$, the active predicate is $\rho^{\Phi_1}(\mathbf{x},t)=\min\big((-\pi/15-\mathbf{x}), (\mathbf{x}+\pi/3), (\pi/3-|\mathbf{x}|)\big)$, and $\forall t\in [12.5,15]$, the active predicate is $\rho^{\Phi_1}(\mathbf{x},t)=\min\big((\pi/3-\mathbf{x}), (\mathbf{x}-\pi/15), (\pi/3-|\mathbf{x}|)\big)$, whereas for the entire time, the active predicate is $\rho^{\Phi_1}(\mathbf{x},t) = \pi/3-|\mathbf{x}|$. We fix the architecture of the neural networks as NCBF: $\{3, \{64\}^3, 1\}$, and NN controller: $\{3, \{64\}^3, 1\}$ with Sigmoid Linear Unit (SiLU) as activation function. We choose a custom layer to represent the cross-coupling terms as \{$e^{-t}\mathbf{x}$\}. The training algorithm converges to obtain the neural networks with $\eta^*=-0.02$, satisfying Theorem \ref{th:4}. In Figure \ref{fig:non_affine}(a)-(c), we show the initial safe set ($\tilde{\mathcal{C}}_{0}^\Phi$) in green color and its refinement at epoch $i=20$ (light blue color), and at epoch $i=40$ (light purple color) while training the neural networks. We observe that the set {$\tilde{\mathcal{C}}_{40}^\Phi(t)\subset \tilde{\mathcal{C}}_{20}^\Phi(t)\subset \tilde{\mathcal{C}}_{0}^\Phi(t)$.} At the end of training, although we have the input constraints, a trajectory starting at the refined safe set $\tilde{\mathcal{C}}_{Epochs}^\Phi(0)$ will be able to remain in the refined safe set $\tilde{\mathcal{C}}^\Phi_{Epochs}(t^{(r)})$.
We also observe in Figure \ref{fig:non_affine}(c) that for different initial states $\mathbf{x}_1(0)=0.5, \mathbf{x}_2(0)=-0.5,\mathbf{x}_3(0)= 0.1$, the trajectories satisfy the specification $\Phi_1$. The control inputs for the three trajectories lie within the safety limits (dotted red lines), as seen from Figure \ref{fig:non_affine}(d). Additionally, the barrier conditions \eqref{cbf_1}, \eqref{cbf_3} are satisfied as seen in Figures \ref{fig:non_affine}(e),(f) with $\alpha=5$. 

\subsection{Mobile robot-Mecanum}
We consider an example of a mobile robot with the following mecanum drive dynamics of the form \eqref{dyn_eq}:
\begin{align}
    \dot{\mathbf{x}}=\begin{bmatrix}
        \dot{\mathbf{x}}_1\\\dot{\mathbf{x}}_2
    \end{bmatrix}=
    \begin{bmatrix}
        \mathbf{u}_1\\\mathbf{u}_2
    \end{bmatrix},
\end{align}
where $\mathbf{x}_1, \mathbf{x}_2$ are the $x,y$ coordinates and $\mathbf{u}_1, \mathbf{u}_2$ are the velocity inputs, bounded within $[-0.2,0.2]$. The neural networks $\mathcal{B}_{\theta_1},\textsl{g}_{\theta_2}$ are trained to satisfy the following STL specification:
\begin{align}
    \Phi_2 =& \square_{[0,15]}(||\mathbf{x}||_2\leq 1.6 \land ||\mathbf{x}-\mathbf{x_u}||_2 > 0.3) \land \square_{[12,15]}(||\mathbf{x}-\mathbf{x_g}||_2\leq 0.3), \label{phi_2}
\end{align}
\begin{figure*}[h!]
    \centering
    \includegraphics[width=\linewidth,trim={0.0cm 0 3.0cm 0},clip]{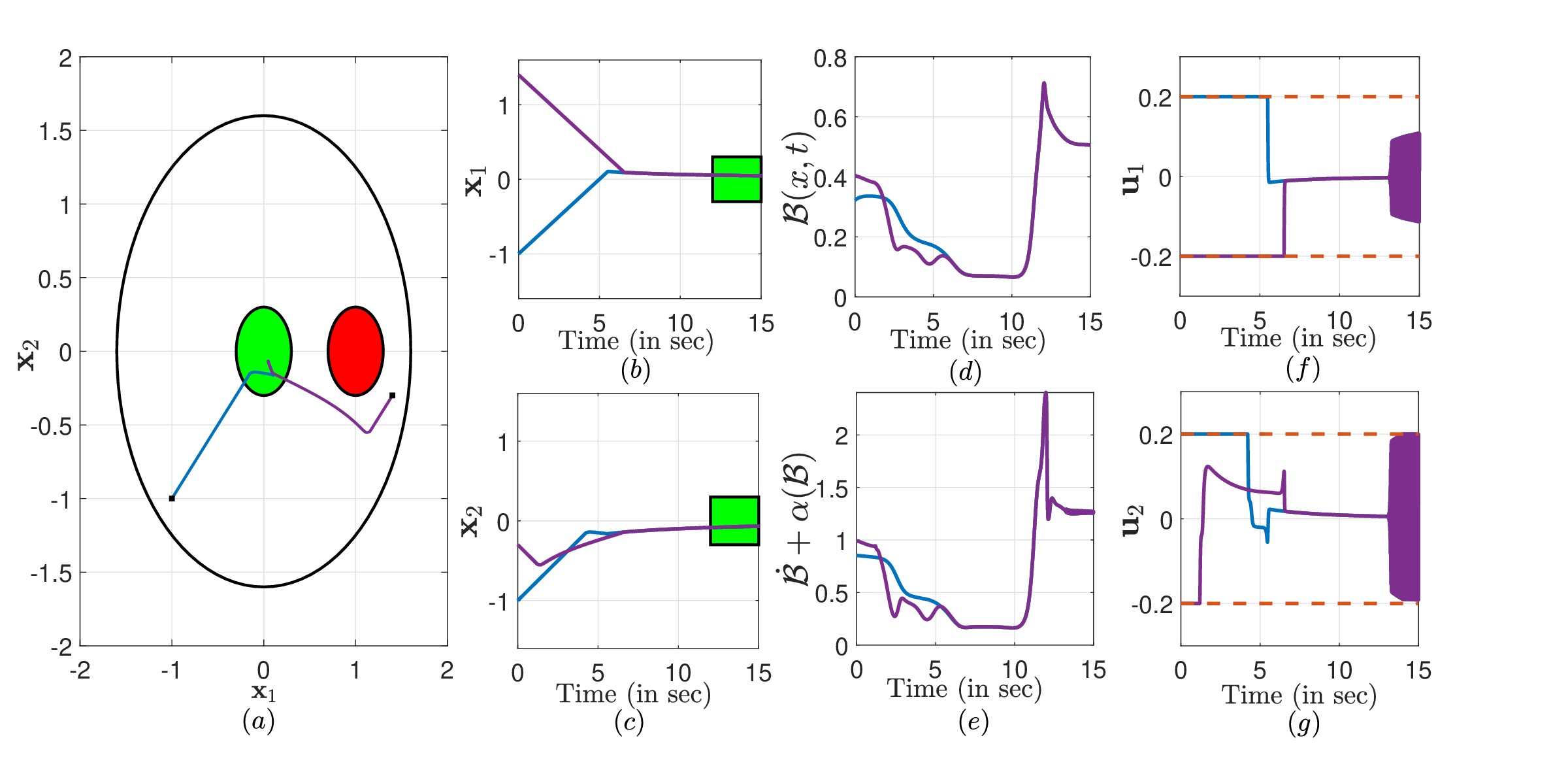}
    \caption{Mecanum satisfying $\Phi_2$ \eqref{phi_2} with state trajectories starting at $[-1,-1]^\top$(blue), $[1.4,-0.3]^\top$(purple), N-TVCBF $\mathcal{B}_{\theta_1}(\mathbf{x}(t),t)$ and controller $\textsl{g}_{\theta_2}(\mathbf{x}(t),t)$ satisfying \eqref{cbf_1}-\eqref{cbf_3}, keeping control inputs within limits (dotted red lines)}
    \label{fig:mecanum}
\end{figure*}
where $\mathbf{x}$ is the robot's position, $\mathbf{x_u} = [1, 0]^\top$ is the unsafe region which the robot should always avoid, and $\mathbf{x_g}=[0,0]^\top$ is the goal position that the robot should reach in the time interval $[12,15]$ seconds. The state space $X = [-2,2]\times[-2,2]$. The discretization parameter for data sampling is $\epsilon=0.02$. 
For every $(x,t)\in W$, the set $\tilde{\mathcal{S}}^{\Phi_2}(t)$ is generated by the predicates formed using the STL $\Phi_2$ (Refer to Section \ref{section_N_TVCBF} and \ref{loss_fun_form}). For example, $\forall t\in [12,15]$, the active predicate is $\rho^{\Phi_2}(\mathbf{x},t)=\min\big((0.3-||\mathbf{x}-\mathbf{x_g}||_2), (1.6-||\mathbf{x}||_2), (||\mathbf{x}-\mathbf{x_u}||_2-0.3)\big)$, whereas for the rest of the time, the active predicate is $\rho^{\Phi_2}(\mathbf{x},t) = \min\big((1.6-||\mathbf{x}||_2), (||\mathbf{x}-\mathbf{x_u}||_2-0.3)\big)$.  We fix the neural network architectures as NCBF: $\{3, \{45\}^3, 1\}$ and NN controller: $\{3, \{45\}^3, 2\}$, with Tanh activation functions in all hidden layers. 
We chose the custom layer to represent the cross-coupling terms as $\{t\mathbf{x}_1, t\mathbf{x}_2\}$. The training algorithm converges to obtain the neural networks with a value of $\eta^*=-0.0664$, satisfying Theorem \ref{th:4}. 
As seen from Figure \ref{fig:mecanum}(a)-(c), we observe that for different initial states $\mathbf{x}(0)=[-1,-1]^\top, \mathbf{x}(0)=[1.4, -0.3]^\top$, the trajectories satisfy the STL specification $\Phi_2$ by reaching the goal position and staying there within the desired time interval $[12,15]$ seconds, while always remaining in safe region. The control inputs for the three trajectories lie within the safety limits (dotted red lines), as seen from Figure \ref{fig:mecanum}(f),(g). Additionally, the barrier conditions \eqref{cbf_1}, \eqref{cbf_3} are satisfied as seen in Figures \ref{fig:mecanum}(d),(e) with $\alpha=2.5$.

\subsection{Pendulum}
\begin{figure*}[h!]
    \centering
    \includegraphics[width=\linewidth,trim={4.0cm 0 3.0cm 0},clip]{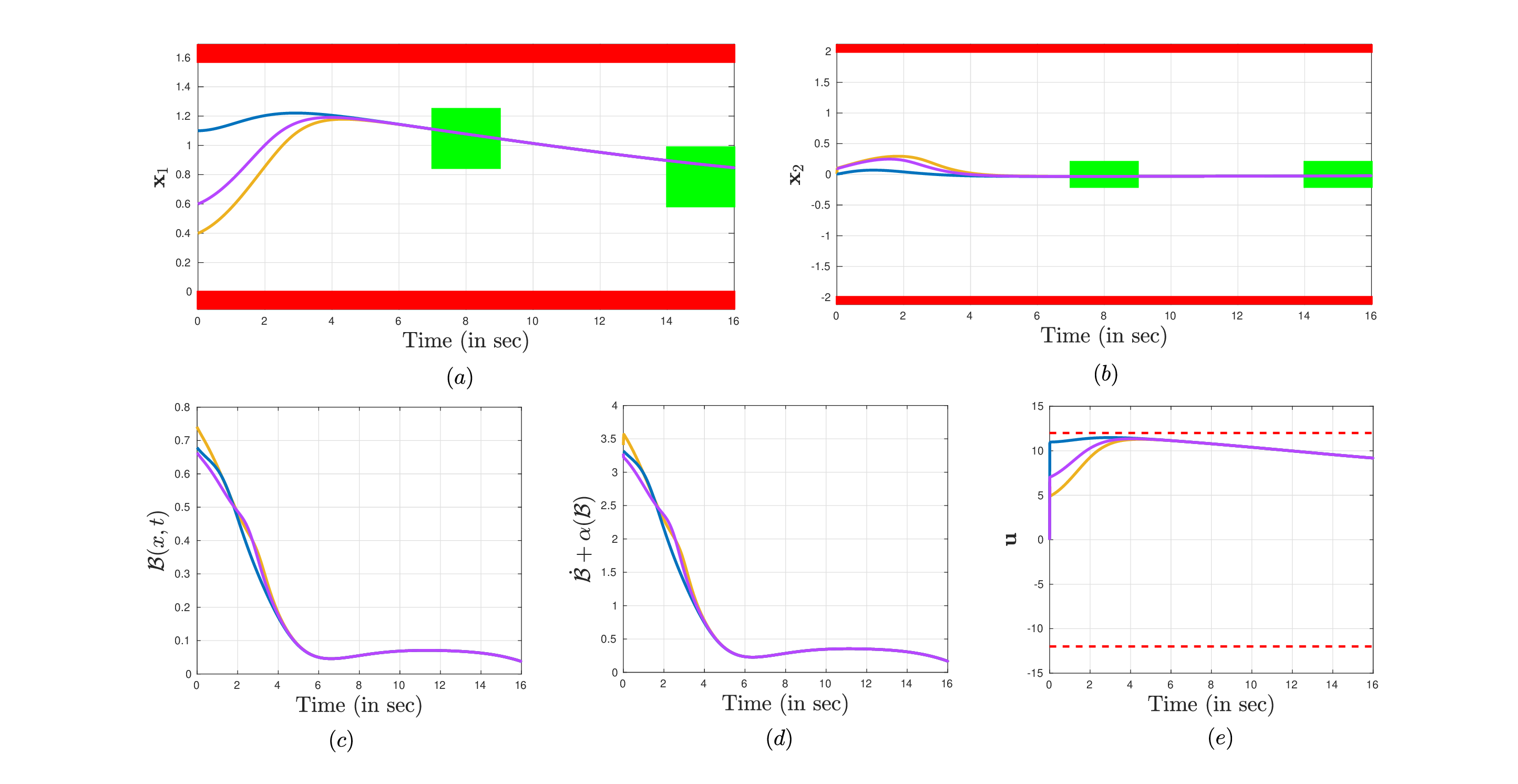}
    \caption{Pendulum satisfying $\Phi_3$ \eqref{phi_3} with state trajectories starting at different initial states $[1.1,0.01]^\top$(blue), $[0.6,0.1]^\top$(purple), $[0.4,0.01]^\top$(yellow), N-TVCBF $\mathcal{B}_{\theta_1}(\mathbf{x}(t),t)$ and controller $\textsl{g}_{\theta_2}(\mathbf{x}(t),t)$ satisfying \eqref{cbf_1}-\eqref{cbf_3}, keeping control inputs within limits (dotted red lines)}
    \label{fig:pendulum_phi_1}
\end{figure*}

We now consider an example of a pendulum with the following dynamics: 
\begin{align}
    \dot{\mathbf{x}}=\begin{bmatrix}
        \dot{\mathbf{x}}_1\\\dot{\mathbf{x}}_2
    \end{bmatrix}=
    \begin{bmatrix}
        \mathbf{x}_2\\
        \frac{\mathbf{u}}{ml^2} - \frac{g}{l}\sin \mathbf{x}_1 - \frac{b}{ml^2}\mathbf{x}_2
    \end{bmatrix},
\end{align}
where $\mathbf{x}_1, \mathbf{x}_2$ are the angle and angular velocity of the pendulum. The mass $m=0.5$kg, length of the rod $l = 0.5$m, acceleration due to gravity $g=9.8$m/s, and the damping coefficient $b=0.1$. The external input $u$ is the torque, bounded between $[-12,12]$. The STL specification for the pendulum considered is as follows: 
\begin{align}
    \Phi_3 = &\square_{[0,16]}(0\leq \mathbf{x}_1\leq\pi/2 \land |\mathbf{x}_2|\leq 2) \land \square_{[7,9]}(|\mathbf{x}_1-\pi/3|\leq 0.2 \land |\mathbf{x}_2|\leq 0.2) \land \nonumber\\& \square_{[14,16]}(|\mathbf{x}_1-\pi/4|\leq 0.2 \land |\mathbf{x}_2|\leq 0.2). \label{phi_3}
\end{align}
To meet the STL specification $\Phi_3$, the pendulum should maintain $\mathbf{x}_1=\pi/3, \mathbf{x}_2=0$ in the time interval $[7,9]$ seconds with a tolerance of $0.2$. Subsequently, in the time interval $[14,16]$, the pendulum should be balanced at the angle $\mathbf{x}_1=\pi/4, \mathbf{x}_2=0$ with a tolerance of 0.2. The state space $X = [-0.15, \pi/2+0.15]\times[-2.15,2.15]$. The states should always remain within $X$.
The discretization parameter for data sampling is $\epsilon=0.06$. For every $(x,t)\in W$, the set $\tilde{\mathcal{S}}^{\Phi_3}(t)$ is generated by the predicates formed using the STL $\Phi_3$ (Refer to Section \ref{section_N_TVCBF} and \ref{loss_fun_form}). We fix the architecture of the neural networks as NCBF: $\{3, \{100\}^5, 1\}$, and NN controller: $\{3, \{128\}^5, 1\}$ with Tanh activation functions in all hidden layers. 
We chose the custom layer to introduce the cross-coupling terms in the form of $\{e^t\mathbf{x}_1, e^t\mathbf{x}_2\}$. The training algorithm converges to obtain $B_{\theta_1}, \textsl{g}_{\theta_2}$ with an optimal value of $\eta^* = -0.04$, satisfying Theorem \ref{th:4}. 
As seen from Figure \ref{fig:pendulum_phi_1} (a) and (b), we observe that the STL specification $\Phi_3$ is satisfied by implementing the trained controller $\textsl{g}_{\theta_3}$ starting at different initial states. The control inputs for the three trajectories lie within the safety limits (dotted red lines), as seen in Figure \ref{fig:pendulum_phi_1}(e). Furthermore, the barrier conditions \eqref{cbf_1}, \eqref{cbf_3} are satisfied as seen from Figures \ref{fig:pendulum_phi_1} (c), (d) with $\alpha=5$.
\begin{figure*}[h]
    \centering
    \includegraphics[width=\linewidth,trim={3.0cm 0 3.0cm 0},clip]{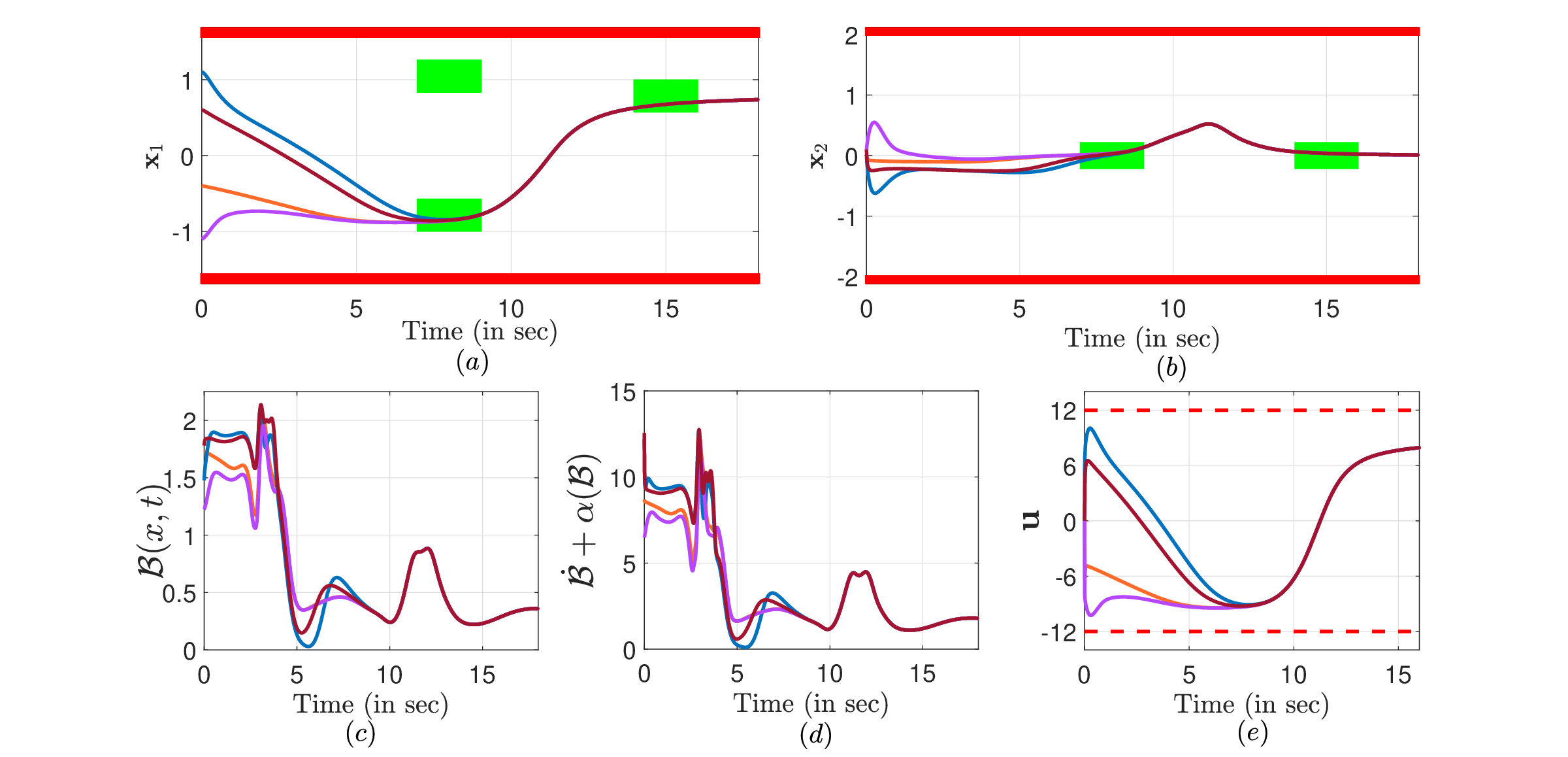}
    \caption{Top row: Pendulum satisfying $\Phi_4$ with state trajectories starting at different initial states represented in blue, brown, purple, orange, Bottom row: N-TVCBF $\mathcal{B}_{\theta_1}(\mathbf{x}(t),t)$ and controller $\textsl{g}_{\theta_2}(\mathbf{x}(t),t)$ satisfying \eqref{cbf_1}-\eqref{cbf_3}, keeping control inputs within limits (dotted red lines)}
    \label{fig:pendulum_phi_2}
\end{figure*}

We consider a different STL specification $\Phi_4$ of the form \eqref{STL_formulae} with a disjunction operator as follows:
\begin{align}
    \Phi_4 = &\square_{[0,16]}(0\leq \mathbf{x}_1\leq\pi/2 \land |\mathbf{x}_2|\leq 2) \land \square_{[7,9]}\big((|\mathbf{x}_1-\pi/3|\leq 0.2  \lor |\mathbf{x}_1+\pi/4|\leq 0.2) \land |\mathbf{x}_2|\leq 0.2 \big) \land \nonumber\\& \square_{[14,16]}(|\mathbf{x}_1-\pi/4|\leq 0.2 \land |\mathbf{x}_2|\leq 0.2). \label{phi_4}
\end{align}
 The state space is $X=[-\pi/2-0.15, \pi/2+0.15]\times [-2.15,2.15]$. To meet the STL specification $\Phi_4$, the pendulum must keep $\mathbf{x}_1$ at $\pi/3$ or $-\pi/4$ and $\mathbf{x}_2$ at 0 with a tolerance of 0.2 in the interval $[7,9]$ seconds, then maintain $\mathbf{x}_1=\pi/4$ and $\mathbf{x}_2=0$ with a tolerance of 0.2 in the interval $[14,16]$, while remaining in $X$ for all $t\in[0,16]$. We use discretization $\epsilon=0.06$ for data sampling. For every $(x,t)\in W$, the set $\tilde{\mathcal{S}}^{\Phi_4}(t)$ is generated using predicates of $\phi_1$ (see Sections \ref{section_N_TVCBF} and \ref{loss_fun_form}). The N-TVCBF and controller networks both uses NN architectures $\{3, \{64\}^3, 1\}$
 with a custom layer for introducing cross-coupling terms of the form $\{e^t\mathbf{x}_1, e^t\mathbf{x}_2\}$  and Tanh activation functions in all hidden layers. With the value of $\eta^*=-0.03$, the training algorithm converges to obtain $B_{\theta_1}, \textsl{g}_{\theta_2}$, satisfying Theorem \ref{th:4}. The Figures \ref{fig:pendulum_phi_2} (a),(b) show that the STL specification $\Phi_4$ is satisfied by implementing the trained controller $\textsl{g}_{\theta_2}$ starting at different initial conditions. The control inputs for the three trajectories lie within the safety limits (dotted red lines), as seen from Figure \ref{fig:pendulum_phi_2}(e).  Furthermore, the barrier conditions \eqref{cbf_1}, \eqref{cbf_3} are satisfied as seen from Figures \ref{fig:pendulum_phi_2} (c),(d), with $\alpha = 5$.
\subsection{Rotating Spacecraft Model}
Consider a rotating rigid spacecraft model \cite{khalil2002nonlinear}, whose dynamics are governed by the following set of equations: 
\begin{align}
\dot{\mathbf{x}}=\begin{bmatrix}
    \dot{\mathbf{x}}_1\\\dot{\mathbf{x}}_2\\\dot{\mathbf{x}}_3
\end{bmatrix}=
\begin{bmatrix}
    \frac{J_2-J_3}{J_1}\mathbf{x}_2\mathbf{x}_3 +\frac{1}{J_1}\mathbf{u}_1\\
    \frac{J_3-J_1}{J_2}\mathbf{x}_1\mathbf{x}_3 +\frac{1}{J_2}\mathbf{u}_2\\
    \frac{J_1-J_2}{J_3}\mathbf{x}_1\mathbf{x}_2 +\frac{1}{J_3}\mathbf{u}_3
\end{bmatrix},
\end{align}
where $\mathbf{x}_1, \mathbf{x}_2, \mathbf{x}_3$ are the angles about the principal axes, the principal moments of inertia are $J_1=200, J_2=200, J_3=100$, and $\mathbf{u}_1, \mathbf{u}_2, \mathbf{u}_3$ are the torque inputs bounded in $[-20,20]$. The state space is $X=[-0.25,0.25]^3$.
\begin{figure*}[h!]
    \centering
    \includegraphics[width=\linewidth,trim={1.0cm 0.5cm 1.0cm 0},clip]{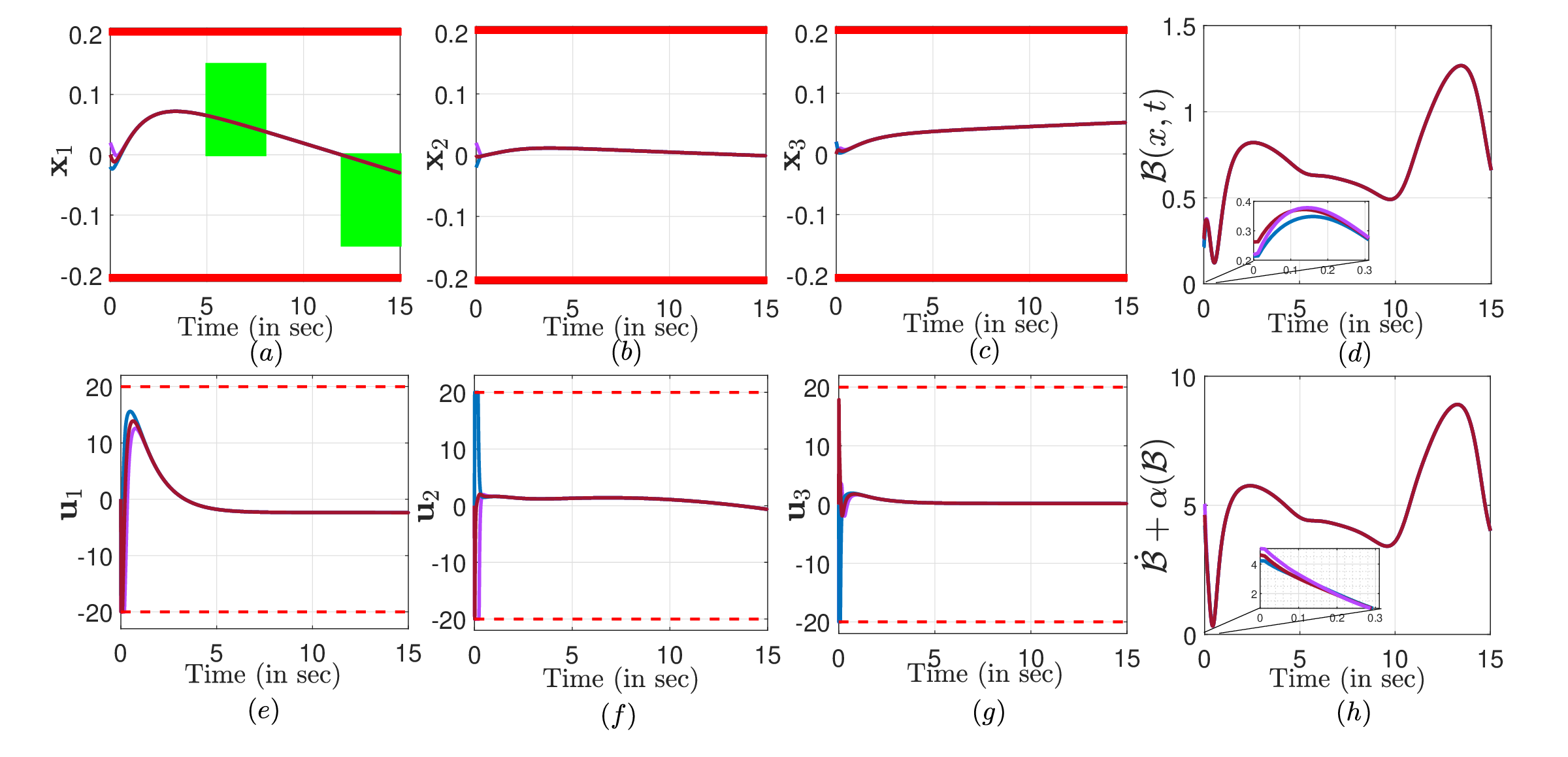}
    \caption{(a)-(c) Spacecraft satisfying $\Phi_5$ with state trajectories starting at different initial states represented in blue, brown, purple, (d)-(h) N-TVCBF $\mathcal{B}_{\theta_1}(\mathbf{x}(t),t)$ and controller $\textsl{g}_{\theta_2}(\mathbf{x}(t),t)$ satisfying \eqref{cbf_1}-\eqref{cbf_3}, keeping control inputs within limits (dotted red lines)}
    \label{fig:spacecraft}
\end{figure*}

We validated the proposed framework on an STL specification:
\begin{align}
    \Phi_5 =&\square_{[0,15]}(||\mathbf{x}||_\infty\leq 0.2) \land \lozenge_{[5,8]}( 0\leq \mathbf{x}_1\leq 0.15)\land \lozenge_{[12,15]}( -0.15\leq \mathbf{x}_1\leq 0) \label{phi5}
\end{align}
The discretization parameter for data sampling is $\epsilon=0.03$. For every $(x,t)\in W$, the set $\tilde{\mathcal{S}}^{\Phi_5}(t)$ is generated by the predicates formed using the STL $\Phi_5$ (Refer to Section \ref{section_N_TVCBF} and \ref{loss_fun_form}).  We fix the architecture of the NNs as N-TVCBF: $\{4, \{50\}^4, 1\}$, and NN controller: $\{4, \{50\}^4, 3\}$ with Tanh activation functions in all hidden layers. 
The custom layer for introducing cross-coupling terms is of the form $\{e^t\mathbf{x}_1, e^t\mathbf{x}_2\}$. The training algorithm converges to obtain $B_{\theta_1}, \textsl{g}_{\theta_2}$ with an value of $\eta^*=-0.03$, satisfying Theorem \ref{th:4}. 
The Figures \ref{fig:spacecraft} (a)-(c) show that the trained controller $\textsl{g}_{\theta_2}$ enforces the STL specification $\Phi_5$ from different initial conditions. The corresponding control inputs remain within the safety limits (dotted red lines) in Figures \ref{fig:spacecraft}(e)–(g). In addition, the barrier conditions \eqref{cbf_1} and \eqref{cbf_3} are satisfied, as shown in Figures \ref{fig:spacecraft}(d) and (h), with $\alpha = 7$.
\section{Conclusion}\label{conclusion}
This study demonstrates the synthesis of a formally verified neural network-based controller that satisfies signal temporal logic (STL) specifications for continuous-time systems. This was achieved by establishing a link between the time-varying control barrier function (TVCBF) and the STL semantics. We formulate the TVCBF constraints as appropriate loss functions for a finite-state space, compute and refine time-varying safe sets for STL satisfaction under input constraints, and, together with a validity condition, provide guarantees for a continuous state space. We also validated the neural network framework for different continuous-time systems subject to fragments of different STL specifications and input constraints.











\bibliographystyle{IEEEtran}
\bibliography{reference}




\end{document}